# Molecular Dynamics Study of Self-Diffusion in Zr

**Mikhail I. Mendelev**[*]

Materials and Engineering Physics, Ames Laboratory, Ames, IA, 50011, USA

**Boris S. Bokstein**

Dept. Physical Chemistry, Moscow State Institute of Steel and Alloys, Moscow, 119049, Russian Federation

[*]mendelev@ameslab.gov

**Abstract**

We employed a recently developed semi-empirical Zr potential to determine the diffusivities in the hcp and bcc Zr via molecular dynamics simulation. The point defect concentration was determined directly from MD simulation rather than from theoretical methods using T=0 calculations. We found that the diffusion proceeds via the interstitial mechanism in the hcp Zr and both the vacancy and interstitial mechanisms give contribution in diffusivity in the bcc Zr. The agreement with the experimental data is excellent for the hcp Zr and for the bcc Zr it is rather good at high temperatures but there is a considerable disagreement at low temperatures.

## 1. Introduction

The atomic mechanism of diffusion in metals has been discussed for many years. It is widely accepted that the vacancy mechanism is the dominant mechanism of self-diffusion in face-centered cubic (fcc) metals[1]. This mechanism can operate because vacancies are an equilibrium defect at any finite temperature and therefore, they are always present in a crystal. The equilibrium vacancy concentration can be found from the following equation

$$x_e^v = e^{-G_f^v/kT} = e^{S_f^v/T} e^{-E_f^v/kT} , \qquad (1)$$

where $S_f^v$, $E_f^v$ and $G_f^v$ are the changes in the entropy, energy and Gibbs energy associated with formation of one vacancy and kT is the thermal factor. Note that in this equation, we neglect the difference between energy and enthalpy for solids. The diffusivity in the vacancy mechanism can be

$$D = a^2 f\gamma v x_e^v e^{-G_m^v/kT} = a^2 f\gamma v x_e^v e^{S_m^v/T} e^{-E_m^v/kT} , \qquad (2)$$

where a is the length of jumps, $\gamma$ is the geometrical factor, f is the correlation factor, $v$ is vibration frequency, $S_m^v$, $E_m^v$ and $G_m^v$ are the entropy, energy and Gibbs energy associated with migration of one vacancy. Finally, combining Eqs. (1) and (2) leads to the following basic equation:

$$D = D_0 e^{-E_D/kT} , \qquad (3)$$

where $D_0$ is the pre-exponential factor which does not depend on temperature and

$$E_D = E_f^v + E_m^v \qquad (4)$$

is the activation energy for self-diffusion. This quantity can be obtained from diffusion experiments which provide the diffusivity as a function of temperature.

In principle, the diffusivity can be obtained from molecular dynamics (MD) simulation but such simulation would require establishing equilibrium vacancy concentration which may require very large simulation cell and rather long simulation time. The formation and migration energies can be rather easily obtained from T=0 calculations (e.g., see [2] or [3]) but using these values at finite temperatures at which the diffusion actually occurs is based on assumption that these quantities do not depend on temperature. The formation and migration energies can be also obtained from more expansive MD simulations where one vacancy is introduced in the simulation cell (e.g., see [4], [5], [6]). The calculation of the formation and migration entropies, correlation factor is more complex and based on further assumptions[2]. Finally, the presence of just one vibration frequency in Eq. (2) is obviously an approximation and in principle, $v$ should be viewed as some average over the vibrational spectrum. Because of complexity of determination of all parameters in Eq. (2), it is seldom used to determine the diffusivity and in the most works, only the activation energy, $E_D$, calculated via Eq. (4) at T=0 is compared with the experimental activation energy.

The vacancy mechanism is also considered as the main mechanism of self-diffusion in hexagonal closely packed (hcp) metals but in this case the dependence of diffusivity on crystallographic orientation should be taken into account. An alternative to the vacancy mechanism of self-diffusion is the interstitial mechanism. In this mechanism the basic equation (Eq. 3) remains unchanged and Eqs. (1) and (4) should be rewritten as follows

$$x_e^i = e^{-G_f^i / kT} = e^{S_f^i / T} e^{-E_f^i / kT} \,, \tag{5}$$

$$E_D = E_f^i + E_m^i \,, \tag{6}$$

where all quantities with superscript "i" are analogous to the corresponding quantities with superscript "v" (e.g., $E_f^i$ is the interstitial formation energy). It should be noted that $S_f^i$ includes not just the change in entropy associated with formation of an interstitial but also a contribution associated with its orientation (if the most stable interstitial is a dumbbell) and a contribution associated with transition of interstitial from one state to another if there are several interstitial configurations with close energies. Usually, the interstitial formation energy is so larger than the vacancy formation energy that the diffusivity according to the interstitial mechanism is much smaller than the diffusivity according to the vacancy mechanism and therefore, the interstitial mechanism does not contribute to the diffusion except of the special case of the radiation damage.

In the case of body-centered cubic (bcc) metals, the interstitial formation energy can be just a little higher than the vacancy formation energy. Since the interstitial migration energy is usually smaller than the vacancy migration energy, the activation energy for diffusion in the interstitial mechanism can be smaller than that in the vacancy mechanism. For example, this result was obtained from the first principles calculations for V in [5]. Strictly speaking, this result does not allow making any final conclusion about the dominant mechanism of diffusion since the pre-exponential factors can be also different in different mechanisms. Therefore, the diffusivity itself rather than the activation energy should be determined for each mechanism to make the final conclusion. In the present study we perform such analysis for pure Zr.

Recently, an embedded atom method (EAM) (see [7]-[8]) potential for Zr was developed (potential #2 in [9]) which properly describes the hcp-bcc phase transformation along with the melting point data. The potential also provides reasonable values for the vacancy and interstitial formation energies in hcp Zr. This makes this potential suitable for the present study. It should be also noted that the first principles calculations show that the bcc Zr is mechanically unstable at T=0. The potential developed in [9] also reproduces this feature and calculations of the vacancy or interstitial formation energies at T=0 lead to negative values which makes impossible to use T=0 calculations for any predictions about self-diffusion in bcc Zr.

The rest of the paper is organized as follows. First, we present the results of MD simulation of the diffusion in a simulation cells with periodic boundary conditions (PBC) in all directions where one point defect is introduced. Next, we present MD simulation of the equilibrium point defect concentration. Finally, we use these data to determine the absolute values of diffusivities and compare obtained results with the experimental data.

**2. Molecular Dynamics Simulation of Self-Diffusion in Zr**

*a) Point defect formation and migration energies*

In this section we present the results of MD simulation of self-diffusion in hcp and bcc Zr. The hcp models contained 2688 atoms and the bcc models contained 2000 atoms in the simulation cell with the periodic boundary conditions in all directions. For both lattices we first ran NVT (constant number of atoms, volume and temperature) MD simulations at several temperatures. At each temperature we determined equilibrium lattice parameters.

For bcc Zr at T=0 the employed FS potential leads to $C_{11}$=96 GPa and $C_{12}$=109 GPa. Thus the bcc phase is mechanically unstable at T=0. To find out the temperature at which the bcc phase becomes mechanically stable we applied a uniaxial strain in z direction keeping the simulation cell size in x and y direction constant and obtained from MD simulation the stress tensor components from which the ratio of the elastic constants was determined as follows:

$$\frac{C_{11}}{C_{12}} = \frac{2\sigma_{zz}}{\sigma_{xx} + \sigma_{yy}} . \qquad (7)$$

The results shown in Fig. 1 demonstrate that bcc Zr becomes mechanically stable above T=200 K. Note that this does not mean that bcc Zr becomes thermodynamically more stable than hcp Zr; the hcp-bcc transformation temperature for the employed potential was estimated as 1233 K in [9].

In order to determine the point defect formation energy we introduced one point defect (vacancy or interstitial) in the simulation cell, performed 20,000 MD steps (1 MD step was equal to ~2 fs) to equilibrate the model and then averaged the energy over the next 2,900,000 MD steps. The point defect formation energy was defined as follows:

$$E_f = E_d - E_p N_d/N, \qquad (8)$$

where $E_p$ is the energy of a perfect system containing N lattice sites and $E_d$ is the energy of the same system with one point defect; $N_d$=N-1 in the case of vacancy and $N_d$=N+1 in the case of interstitial. Since the simulation cell had periodic boundary conditions in all directions and contained only single crystal no additional point defects could arise during MD simulation unless vacancy – interstitial pairs formed (see Section 2b). The obtained point defect formation energies are shown in Fig. 2. In the case of the hcp lattice, the vacancy formation energy slightly increases in the temperature interval from 0 to ~1400 K and then the increase becomes much more pronounced. A similar temperatures dependence of the vacancy formation energy was observed for fcc metals (Al in [10] and Cu in [6]). The interstitial formation energy also slightly increases in the temperature interval from 0 to ~1200 K but then decreases. A more complicated temperature dependence of the interstitial formation energy can be related to the fact that there are several possible interstitial types in hcp Zr (e.g., see [11]) and interstitial types which never occur at low temperatures give contribution at high temperatures. This issue deserves a special consideration. In the case of the bcc lattice, both the vacancy and interstitial formation energies increases in the temperature interval from 0 to ~1800 K and then dramatically drop with increasing temperature. The reason for this drop will be discussed in the next section.

The same series of MD simulations were used to determine the diffusivities associated with two types of point defects. We define the effective diffusivity as

$$D^{eff} = D^{sim}/x^d , \qquad (9)$$

where $x^d$ is the point defect concentration in the simulation cell and $D^{sim}$ is determined from the dependence of the mean square displacement of atoms $<\Delta r^2>$ on the simulation time, t:

$$D^{sim} = \frac{<\Delta r^2>}{6t} \ . \qquad (10)$$

The effective diffusivity is different from the actual diffusivity because the concentration of the point defects in these series was $1/N_m$, where $N_m$ is the number of atoms in the simulation cell, rather than the equilibrium point defect concentration. If atoms can diffuse only via exchange with point defects and the number of point defects is fixed then sum of square displacements of atoms does not depend on $N_m$ and $<\Delta r^2>$ is inversely proportional to $N_m$. Therefore, the effective diffusivity does not depend on the number of atoms in the simulation cell. It should be noted that since the point defect concentration in the simulation was the same at all temperatures the activation energy for the effective diffusivity is equal to the point defect migration energy:

$$D^{eff} = D_0^{eff} e^{-E_m/kT} \ . \qquad (11)$$

The effective diffusivities are shown in Fig. 3 in the Arrhenius coordinates. The data for the hcp fall on the straight lines up to ~1800 K. At higher temperatures small positive deviations can be seen. The data for the bcc lattice fall on straight lines up to ~1700 K. At higher temperatures very strong positive deviations can be seen. Note that about the same temperature, the point defect formation energies in the bcc lattice dramatically drop.

The point defect migration energies and the pre-exponential factor $D^{eff}$ obtained from the data shown in Fig. 3 are presented in Table I. For both lattices, the migration energy in the interstitial mechanism is much smaller than that in the vacancy mechanism such that the sum of the interstitial formation and migration energies (the activation energy for diffusion in the interstitial mechanism) is smaller than that in the vacancy mechanism. The activation energies for both lattices are in good agreement with experimental data.

*b) Formation of vacancy – interstitial pairs*

In this section we discuss the reason for the dramatic increase of the diffusion and drop in point defect formation energies at high temperatures in the bcc lattice. Both effects can be easily explained if we assume that at high temperatures vacancy-interstitial pairs can spontaneously form. It would obviously accelerate the diffusion because of the increase of the point defect concentration in the simulation cell. It would also lead to a decrease in the point defect formation energy since it will lead to an increase of the value of $E_p$ calculated using simulation cells containing "perfect" crystals which in reality can already have defects. Indeed checking of the models of "perfect" bcc crystal showed that at high temperatures they always have some defects which lead to non-zero diffusivity in these models [9].

It should be noted that the probability of formation of the vacancy – interstitial pairs cannot be determined from T=0 calculations since both the vacancy and interstitial formation energies are negative for the bcc phase. However, obviously if we observe the formations of such pairs even in a simulation cell containing only 2000 atoms, we can obtain a reasonable statistics from larger models. In order to get such a statistics we performed new series of MD runs where bcc models contained 27648 atoms and the simulation time was ~5.9 ns (3,000,000 MD steps). At T=1800 K, the simulation contains only one or two vacancy – interstitial pairs. Therefore, we could not get any reliable statistics at lower temperatures. As can be seen in Fig. 4, starting from ~1900 K the data almost exactly fall on a straight line in the Arrhenius coordinates.

In reality, when the bulk part of the system is surrounded by some kind of interface, the point defect can come from or go out through this interface. The vacancy and interstitial concentrations are independent and determined by corresponding formation entropy and enthalpy. In the case of simulation with periodic boundary conditions, the vacancy and interstitials can form only as pairs and their concentration is always the same. It is easy to derive an expression for this concentration following the classical derivation for the equilibrium vacancy concentration (e.g., see [1]). Consider simulation cell with periodic boundary conditions in all directions containing $N_s$ sites. The change of the Gibbs free energy upon formation of $N_p$ vacancy – interstitial pairs is

$$G_f = N_p G_f^p - TS_c , \qquad (12)$$

where $G_f^p$ is the change of the Gibbs free energy associated with formation of one pair and $S_c$ is the change of entropy associated with different possibilities to arrange vacancies and interstitial within the simulation cell. This change of entropy can be written as

$$S_c = k \ln \frac{N_s!}{(N_s - 2N_p)!(N_p!)^2} . \qquad (13)$$

Inserting this expression in Eq. (12) and using the Stirling's approximation we find

$$G_f = N_p G_f^p + kT \left\{ (N_s - 2N_p) \ln \frac{N_s - 2N_p}{N_s} + 2N_p \ln \frac{N_p}{N_s} \right\} \qquad (14)$$

The equilibrium concentration of the vacancy – interstitial pairs can be found from the condition $dG_f/dN_p=0$, which yields:

$$x_e^p = \frac{N_p}{N_s} = e^{-G_f^p / 2kT} = e^{S_f^p / 2k} e^{-E_f^p / 2kT} . \qquad (15)$$

Since $G_f^p = G_f^v + G_f^i$, the equilibrium vacancy – interstitial pair concentration in a simulation with PBC in all directions can be expressed as

$$x_e^p = \sqrt{x_e^v x_e^i} , \qquad (16)$$

where $x_e^v$ and $x_e^i$ are equilibrium vacancy and interstitial concentrations in a system with interfaces ($x_e^v \neq x_e^i$!).

The obtained expression explains why the data fall on a straight line in the Arrhenius coordinates. The energy of the formation of a vacancy – interstitial pair found from the data shown in Fig. 4 is equal to $E_f^p$=5.25 eV/atom. This quantity should be equal to the sum of the vacancy and interstitial formation energies. The sum of the vacancy and interstitial formation energies shown in Fig. 2 never reaches 5.25 eV/atom. However, as was discussed above we cannot trust our MD simulation data for the point defect formation energies above ~1700 K while the vacancy – interstitial pair concentrations were obtained above 1900 K. Perhaps, at high temperature the point defect energies increase much faster with increasing temperature as can be seen for the hcp vacancy formation energy in Fig. 2.

We also performed similar MD runs for the hcp lattice. In this case the simulation cell contained 27540 atoms and the simulation time was 2,900,000 MD steps. While hcp phase has lower melting temperature than the bcc phase (1913 K vs. 2109 K [9]) we observed formation of the vacancy – interstitial pairs only starting from T=1900 K. Their concentration was almost 6 times smaller than that in the bcc phase at the same temperature. taking into account that the sum of the vacancy and interstitial formation energies in the hcp phase is much higher than that in the bcc phase (see Fig. 2), there should not be any effect of the spontaneous formation of the vacancy – interstitial pairs on the MD results below T=1900 K. Above this temperature, some acceleration of diffusion in the hcp phase can be indeed seen in Fig. 3 (although not so pronounced as in the bcc phase).

*c) Point defect concentration*

The MD simulations described in the previous section utilized simulation cells with the periodic boundary conditions in all directions. Therefore, vacancies and interstitials could form only as pairs and their concentration was always the same. In reality both vacancies and interstitials can come from or go out through interfaces and their concentrations are not the same. We also noted that our MD data for the point defect formation energies are reliable only below 1700 K. In order to find the point defect concentrations and their formation energies at high temperatures we performed new series of MD simulations where the simulation cell had periodic boundary conditions in x and y directions and free surfaces in z directions. The hcp models in this series contained 27,200 atoms and the simulation time was ~30 ns (15,000,000 MD steps). The bcc models contained 27,000 atoms and the simulation time was ~20 ns (10,000,000 MD steps). The distance between two free surfaces was ~210-220 Å.

The results for the bcc phase are shown in Fig. 4. The vacancy concentration is larger than the vacancy – interstitial pair concentration obtained in the simulation series with PBC in all directions and the interstitial concentration is smaller. The geometrical mean of the vacancy and interstitial concentration coincides with $x_e^p$ in full accordance with Eq. (16) which indicates that in both cases the simulation time was sufficient to reach equilibrium.

As further test that the simulation time was sufficient to reach equilibrium point defect concentration we performed two additional series of MD simulations. In the first series we formed 20 vacancies within the bulk part of the simulation sell and in the second series we formed 40 vacancies within the bulk part of the simulation sell. As can be seen from Fig. 3, vacancies diffuse slower then interstitials therefore, if the simulation time is sufficient to establish the equilibrium vacancy concentration it should be definitely sufficient to establish the equilibrium interstitial concentration. Figure 5 shows the vacancy concentration as function of the simulation time at T=1800 K. The examination of this figure shows that even in the case of 40 vacancies in the initial simulation cell, 14 ns is sufficient to reach the equilibrium vacancy concentration. Obviously, at higher temperatures this time should be even smaller.

The vacancy formation energy obtained from equilibrium vacancy concentration is 2.34 eV/ atom and the interstitial formation energy obtained from the equilibrium interstitial concentration is 2.68 eV/atom. Both values are a little higher the values shown in Fig. 2 which indicates that at high temperature the formation energies increase faster than it can be predicted by extrapolation of low temperature data.

The equilibrium point defect concentrations in the hcp phase are shown in Fig. 6 as function of temperature. In contrast to the bcc lattice, the vacancy and interstitial concentrations are rather similar. In the hcp phase, both concentrations decrease with decreasing temperature much faster than corresponding concentrations in the bcc phase. Because of lower defect concentrations in the hcp phase, the statistical error of the MD data in this case is higher than that in the bcc phase.

Summarizing results of this section and Section 2a we conclude that MD simulation gives reliable data on the point defect formation energy at low temperatures until vacancy-interstitial pairs start spontaneously forming during MD simulation. On the other hand, MD simulation provides reliable point defect concentration data at high temperatures while at low temperatures no reasonable statistics can be obtained. In order to calculate the actual diffusivity and compare it with experimental data we need to obtain the point defect concentration at temperatures much lower than we used in MD simulation of the equilibrium point defect concentration. In order to derive an expression for these temperature dependences we assume that the point defect formation free energy takes the following form:

$$G_f = g_1 + g_2 T + g_3 T \ln T + g_4 T^2 . \tag{17}$$

The point defect formation energy can be derived via the Gibbs-Helmholtz equation:

$$\frac{\partial (G_f / T)}{\partial T} = -\frac{E_f}{T^2} , \tag{18}$$

which yields:

$$E_f = g_1 - g_3 T - g_4 T^2 . \tag{19}$$

Finally the formation entropy can be obtained via relationship $S_f = -\partial G_f / \partial T$, which yields:

$$S_f = -g_2 - g_3 - g_3 \ln T - 2 g_4 T . \tag{20}$$

The coefficients $g_k$ can be found from fitting to the data on the formation energy at low temperatures and the point defect equilibrium concentration at high temperatures, x (recall that $G_f = -kT \ln x$). The obtained coefficients are presented in Table II. Figures 2 and 6 demonstrate that the obtained interpolation expressions provide good agreement with the simulation data on the point defect formation energies at low temperatures and equilibrium concentrations at high temperatures.

Figure 7 shows the point defect formation entropy. The values themselves are reasonable comparing to the results of T=0 calculations performed for other metals within harmonic approximation (e.g., see [12]). However, the entropies in Fig. 7 depend on temperature which may be result of anharmonic vibrations. The point defect entropies for the bcc phase become negative at low temperatures which may reflect the fact that the bcc lattice is mechanically unstable at these temperatures. However, it should be noted that this is rather extended extrapolation from the high temperature which may be not very accurate itself. It should be also noted that the chosen form of the free energy function (Eq. 17) can also affect the results. For example, this form cannot be valid at very low temperatures.

*d) Self-diffusion coefficients*

Once the effective diffusivity and point defect concentration have been determined, the diffusivity can be calculated as
$$D = x_e D^{eff}. \tag{21}$$
The results are shown in Figs. 8 and 9. The examination of Fig. 8 shows that the self-diffusion in the hcp Zr proceeds via the interstitial mechanism. The simulation results are in the remarkable agreement with the experimental data. The examination of Fig. 9 shows that both vacancy and interstitial mechanisms in the bcc Zr lead to about the same diffusivities. The point is that the faster interstitial diffusion is compensated by their smaller concentration. The experimental data for the bcc Zr are also shown in Fig. 9. It should be noted that since both interstitial and vacancy mechanism operate at the same time the sum of diffusivities by both mechanisms should be compared with the experimental data. The agreement at high temperature is satisfactory. The agreement at lower temperatures is much worse such that the difference between the simulation and experimental data reaches almost 2 orders of magnitude at 1200 K (~0.56$T_m$).

## 3. Discussion and concluding remarks

In this work we used molecular dynamics simulation to determine the diffusivities in the hcp and bcc Zr. In both cases the point defect concentration was determined directly from MD simulation rather than from theoretical methods using T=0 calculations (which would be simply impossible to perform for the bcc Zr). The agreement with the experimental data is very good for the hcp Zr and for the bcc Zr there is considerable disagreement at low temperatures. In particular, this disagreement can be attributed to the quality of the employed EAM potential. Indeed, the employed potential was fit only to the vacancy formation energy in the hcp Zr. It provides the right interstitial formation energy for the hcp Zr[9] (although it makes the basal octahedral interstitial to be the most stable, while the first principles calculations show that the octahedral interstitial has slightly lower energy[11]). The potential was not fit to the bcc point defect formation energies since this phase is mechanically unstable at T=0 and only unrelaxed point defect formation energies can be obtained at T=0 which have no physical sense especially in the case of interstitials.

The simulation allowed establishing several important results. First, it showed that both parts of diffusion activation free energy, activation energy and activation entropy, depend on temperature. It should be noted that the values of the activation entropy, which is usually considered to be very small or even negative (e.g., see the analysis of experimental data in [13]) are sufficiently large, especially at high temperatures. For example, at 2000K the vacancy formation entropy in the bcc Zr is $S_f^v/k = 5.4$ which gives factor $\exp(S_f^v/k) = 220$ in the diffusivity in the vacancy mechanism. This high vacancy formation entropy provides that the vacancy mechanism gives about the same contribution to the diffusivity as the interstitial mechanism while the activation energy for diffusion is smaller in the interstitial mechanism. In the hcp Zr, the activation energy for the interstitial mechanism is considerably smaller than that in the vacancy mechanism while the interstitial concentration is about the same as the vacancy concentration. Thus the interstitial mechanism is the dominant mechanism of diffusion in the hcp Zr.

In the case of the bcc Zr, we could not explain the well known anomalous behavior of Zr (and also Ti, U, possibly, V [14]). For the bcc Zr, the experimental dependence of lnD on 1/T is not linear (Fig. 9) [15]. In the temperature range above the temperature of hcp-bcc phase transformation (from 1136K to approximately 1700 – 1800K) the activation energy and the preexponential factor for self-diffusion are well below than it follows from the commonly accepted correlations: $E_D=18RT_m$ , $D_0 = 10^{-5}$ $m^2/s$ [1]. The simulation also predicts non-linear lnD vs. 1/T dependence which is related to the temperature dependence of the point defect formation energies in the bcc Zr. However, the disagreement with the experimental data at low temperatures is too large. Therefore, other possibilities to explain the experimental data should be considered. For example, decreasing in diffusivity comparing to the Arrhenius law could be explained by high density of dislocations which are generated at phase transformations and remain stable up to high temperatures [16] or high excess point defect concentration associated with solute, probably, with oxygen [17].

In equilibrium the system contains an equilibrium amount of point defects and the mechanism of formation of these defects does not play any role. However, if the system is not in equilibrium, the time which is necessary to establish the equilibrium point defect concentration and, therefore, the point defect concentration itself do depend on the mechanisms of the defect formation. From this point of view, it is very important that in the case of bcc Zr, it is not necessary for vacancies to move to/out a surface or interface (grain boundary, phase boundary, etc.) for disappearance/appearance. Instead, a vacancy can annihilate with an interstitial from the vacancy – interstitial pair appeared around while the vacancy from this pair will continue to migrate. This mechanism can considerably accelerate the vacancy migration. This can be extremely important under a radiation damage.

**Acknowledgements**
Work at the Ames Laboratory was supported by the Department of Energy, Office of Basic Energy Sciences, under Contract No. DE-AC02-07CH11358. Work at MISIS was supported by the Ministry of Education under the contract No. 02.513.11.3402 and by RFBR under grant No. 08.03.00498.

Table I. The diffusion parameters. The defect formation and migration energies taken from MD runs at 1200K.

| Phase | Defect | $D_0^{eff}$ (cm$^2$/s) | $E_f$ (eV/atom) at T=1200 K | $E_m$ (eV/atom) | $E_f + E_m$ (eV/atom) | $E_D$ (experiment) (eV/atom) |
|---|---|---|---|---|---|---|
| hcp | vacancy | 6.11·10$^{-2}$ | 2.41 | 1.20 | 3.61 | 3.17[18] |
|  | interstitial | 6.03·10$^{-4}$ | 3.04 | 0.14 | 3.18 |  |
| bcc | vacancy | 1.53·10$^{-3}$ | 1.88 | 0.46 | 2.34 |  |
|  | interstitial | 5.21·10$^{-4}$ | 2.05 | 0.11 | 2.16 | 2.04[15] |

Table II. The coefficients of the interpolation expressions for the point defect formation Gibbs energies (Eq. 15).

| Lattice | Defect | $g_1$ | $g_2$ | $g_3$ | $g_4$ |
|---|---|---|---|---|---|
| bcc | vacancy | 1.517 | 1.55·10$^{-3}$ | -1.92·10$^{-4}$ | -9.01·10$^{-8}$ |
|  | interstitial | 1.811 | 8.40·10$^{-4}$ | -1.15·10$^{-4}$ | -3.87·10$^{-8}$ |
| hcp | vacancy | 2.309 | -1.74·10$^{-3}$ | 2.61·10$^{-4}$ | -2.77·10$^{-7}$ |
|  | interstitial | 2.934 | -4.51·10$^{-4}$ | -2.34·10$^{-5}$ | - |

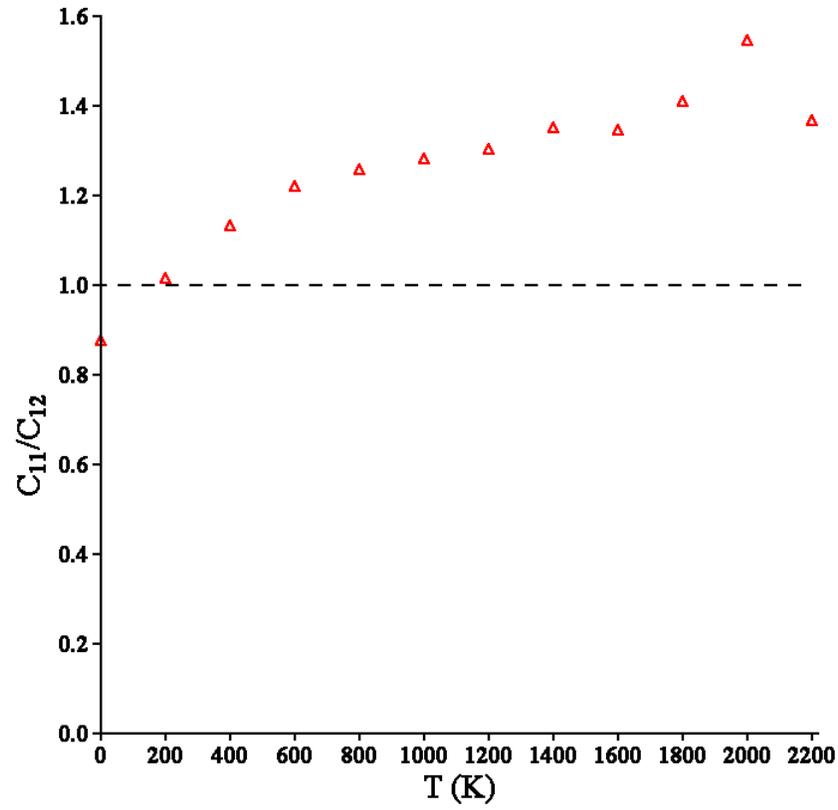

Figure 1. $C_{11}/C_{12}$ ratio for bcc Zr.

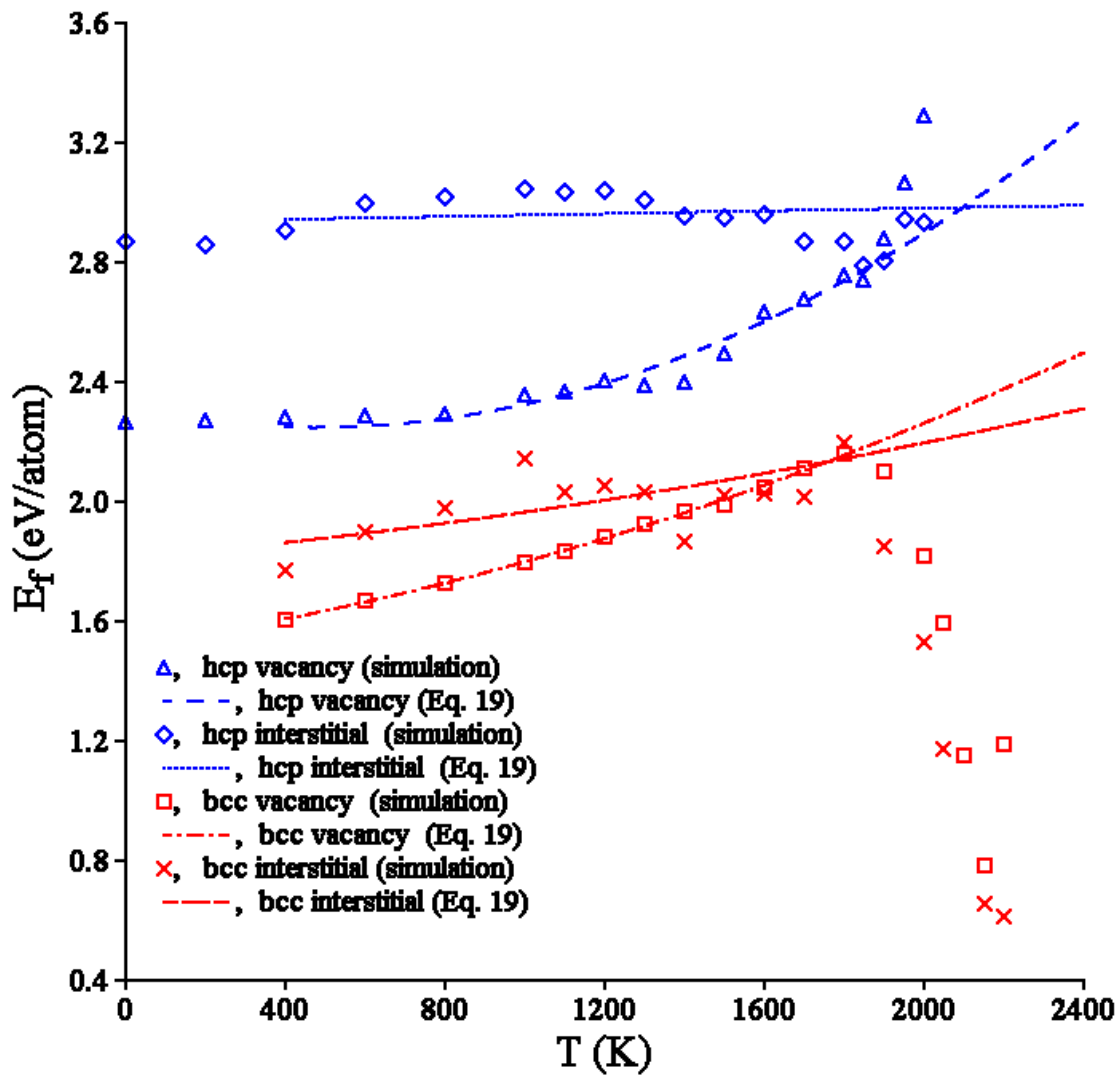

Figure 2. Point defect formation energy as function of temperature.

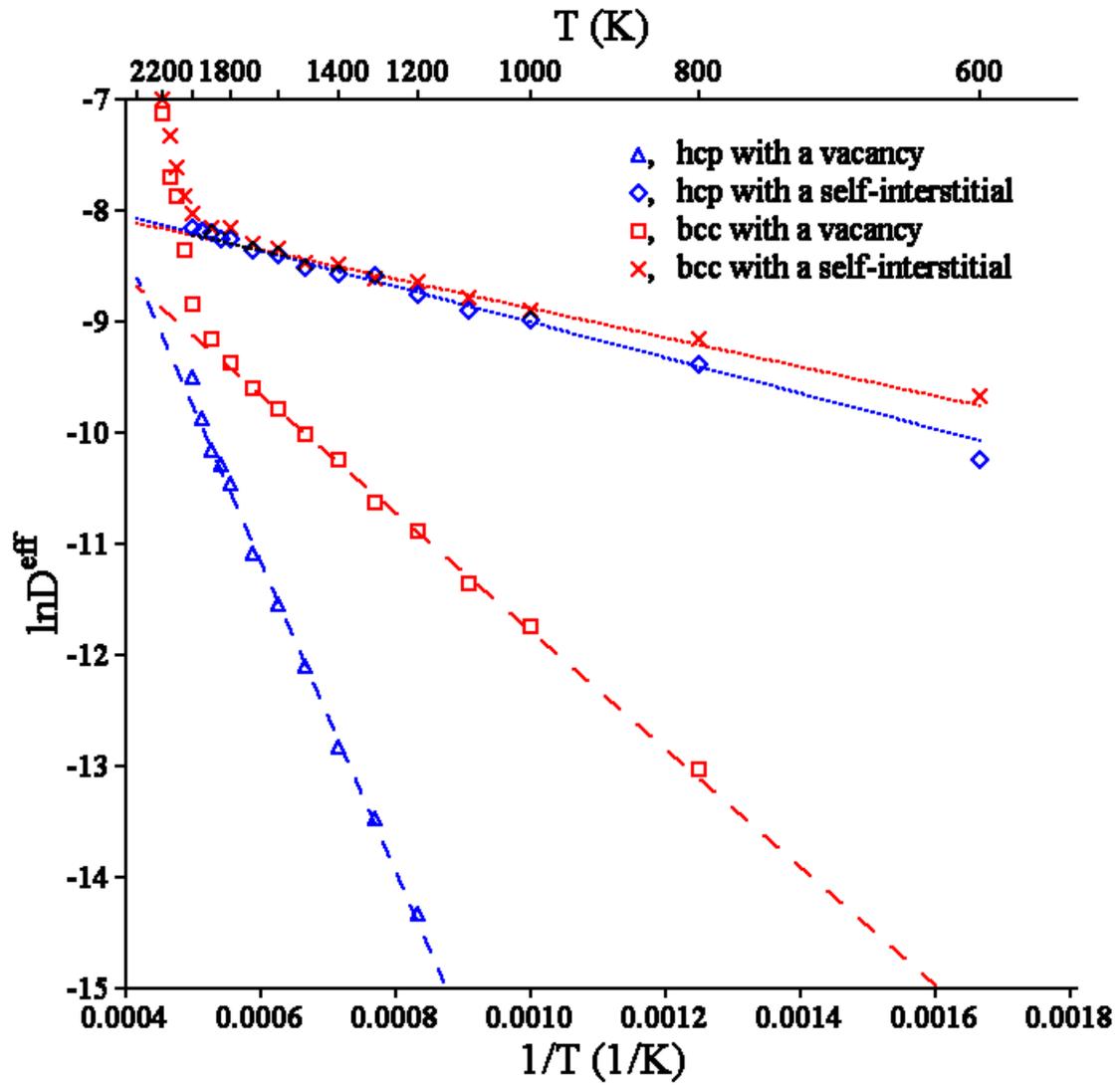

Figure 3. The effective diffusivity as function of temperature. The straight lines are created using the activation energies and preexponential factors presented in Table I.

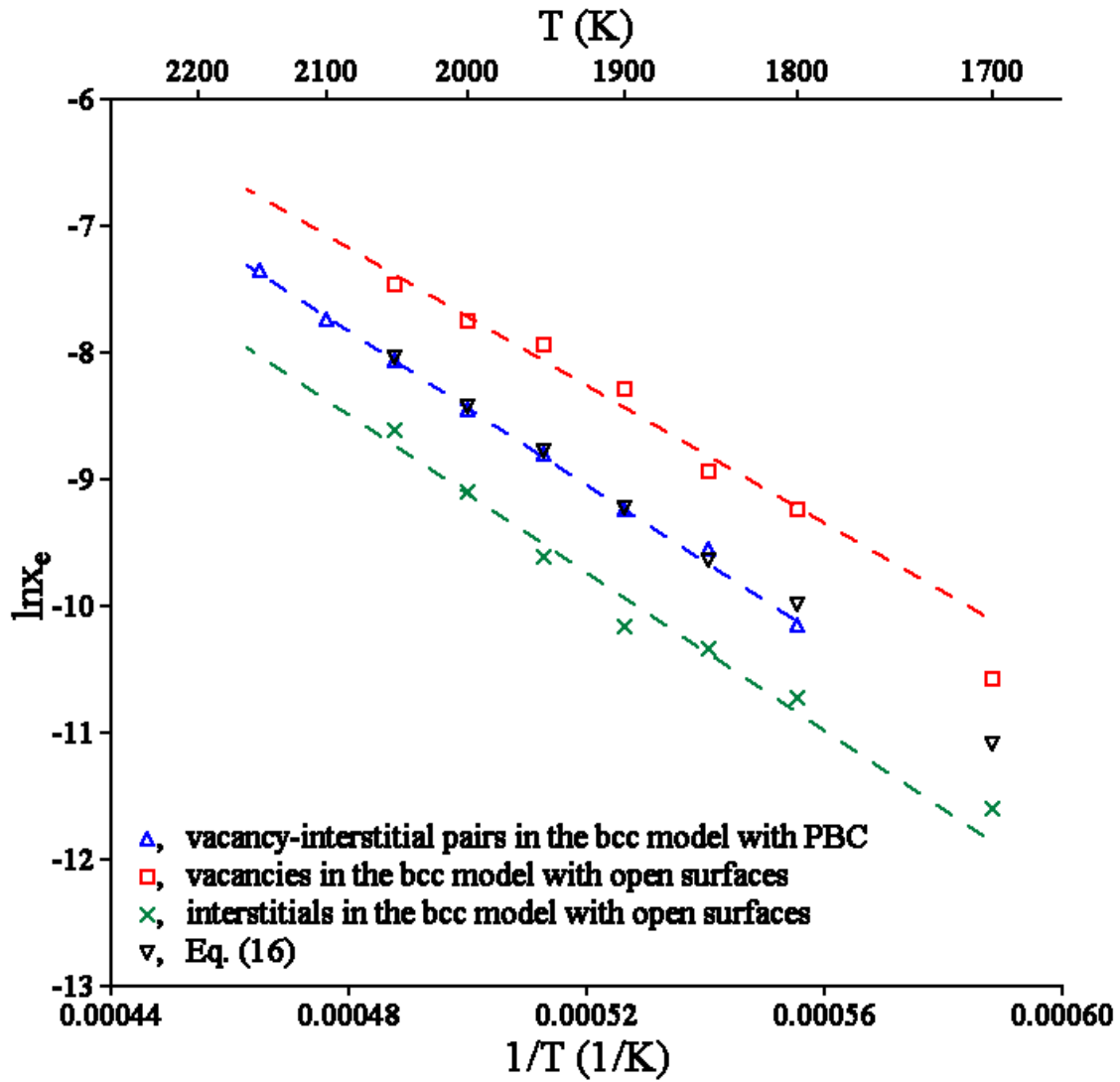

Figure 4. Point defect concentration as function of temperature.

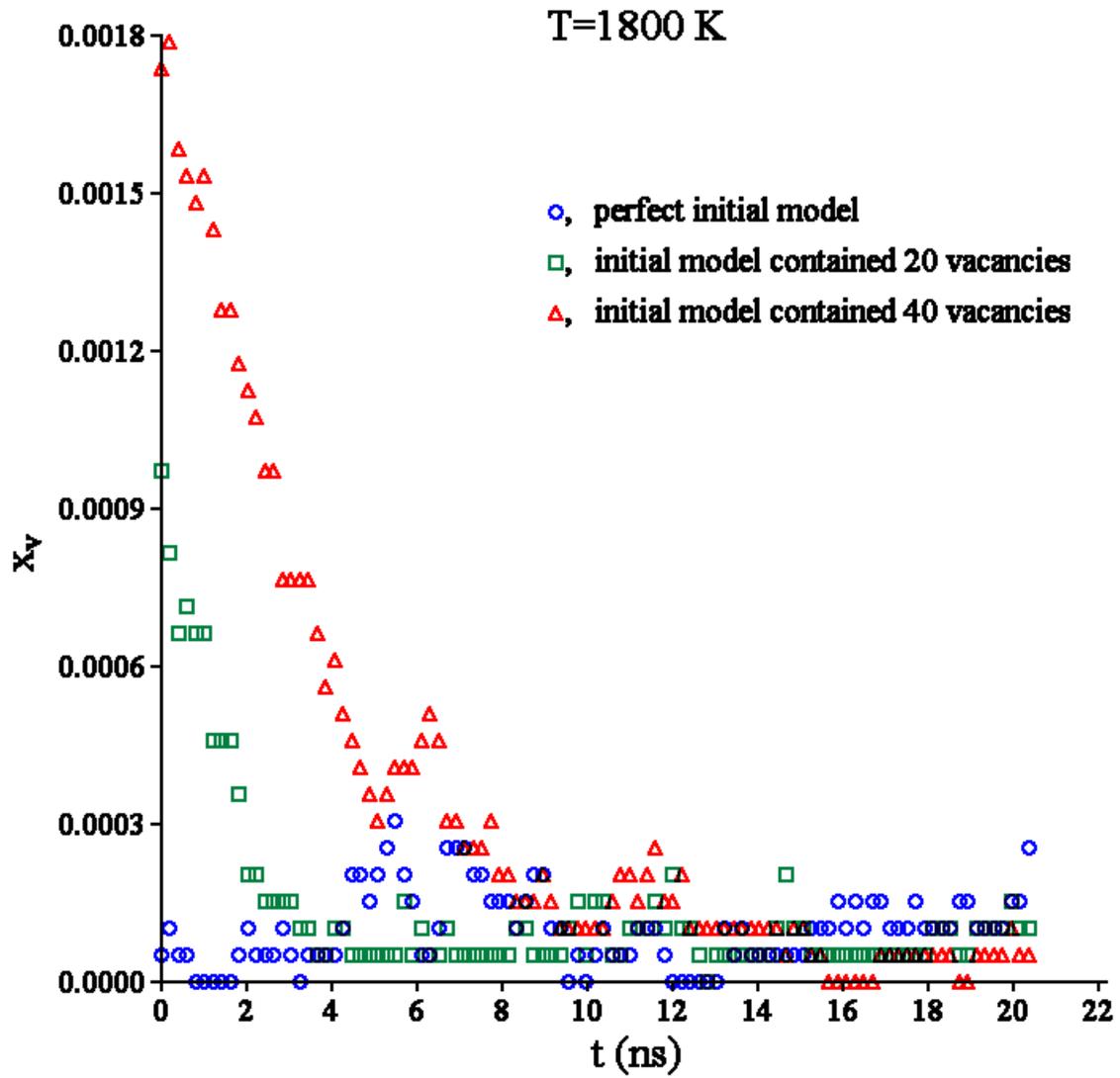

Figure 5. The vacancy concentration as function of the simulation time.

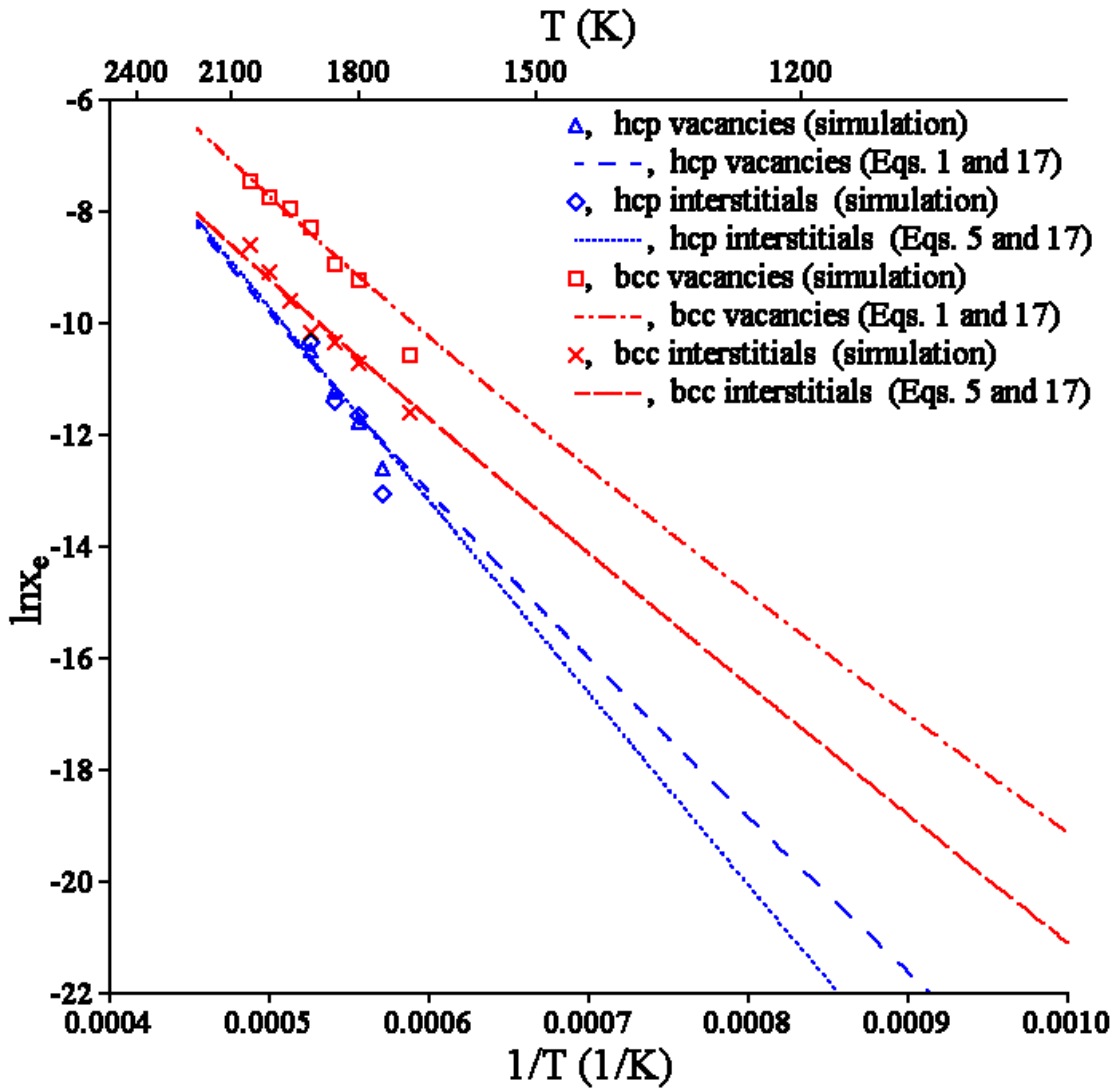

Figure 6. Point defect concentration as function of temperature.

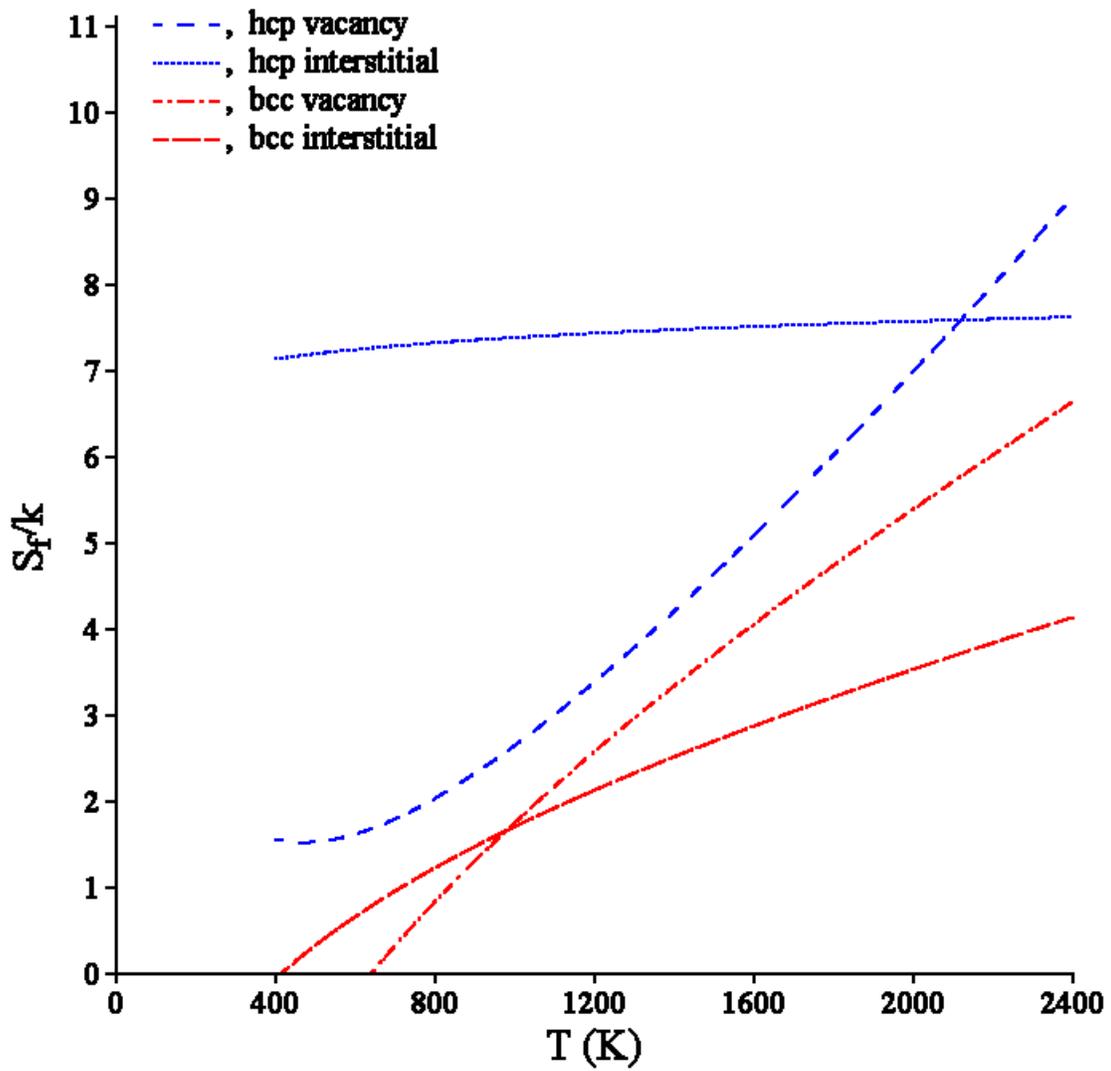

Figure 7. Point defect formation entropy as function of temperature.

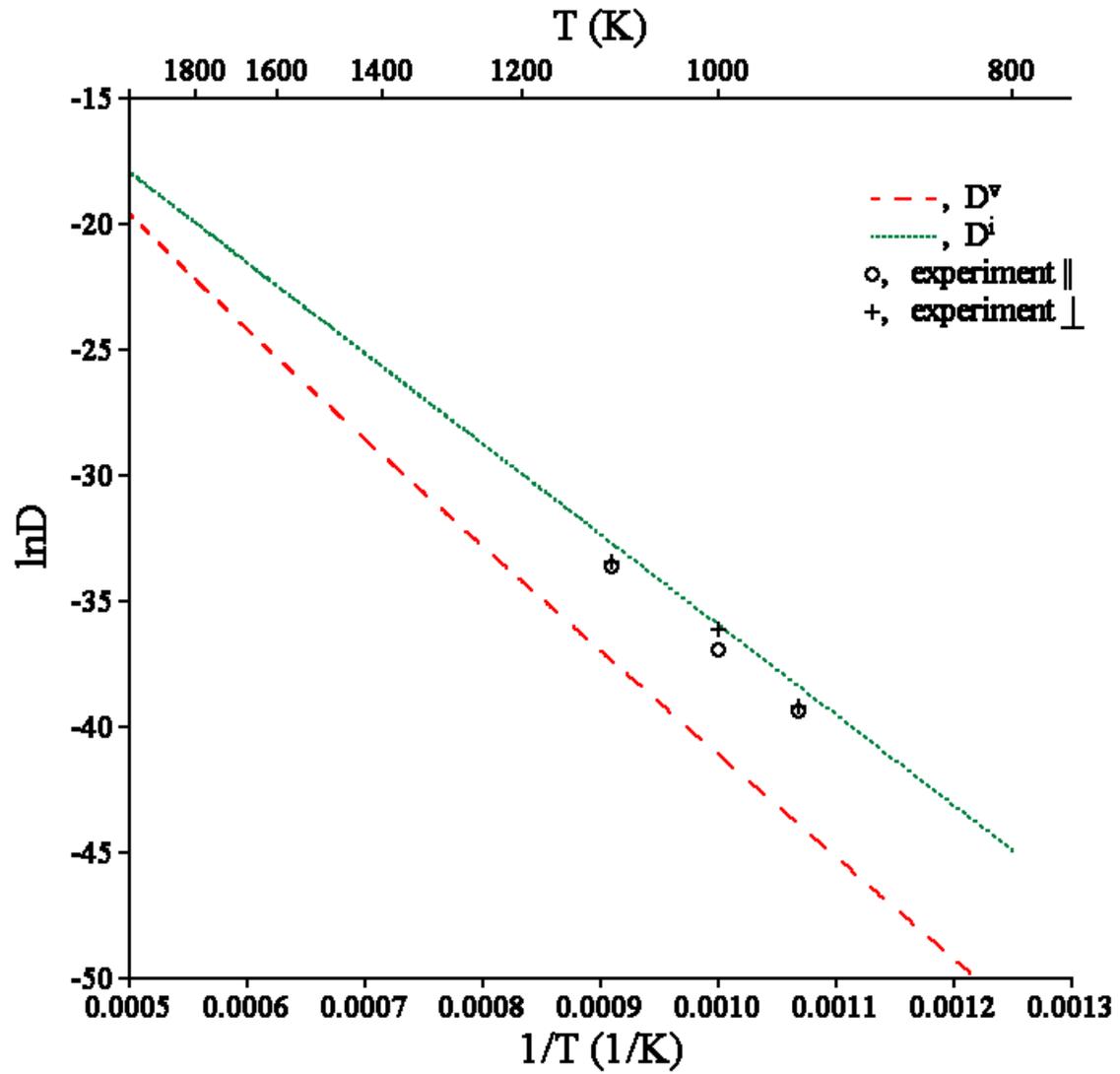

Figure 8. Diffusivity in hcp as function of temperature. The experimental data are from [18]

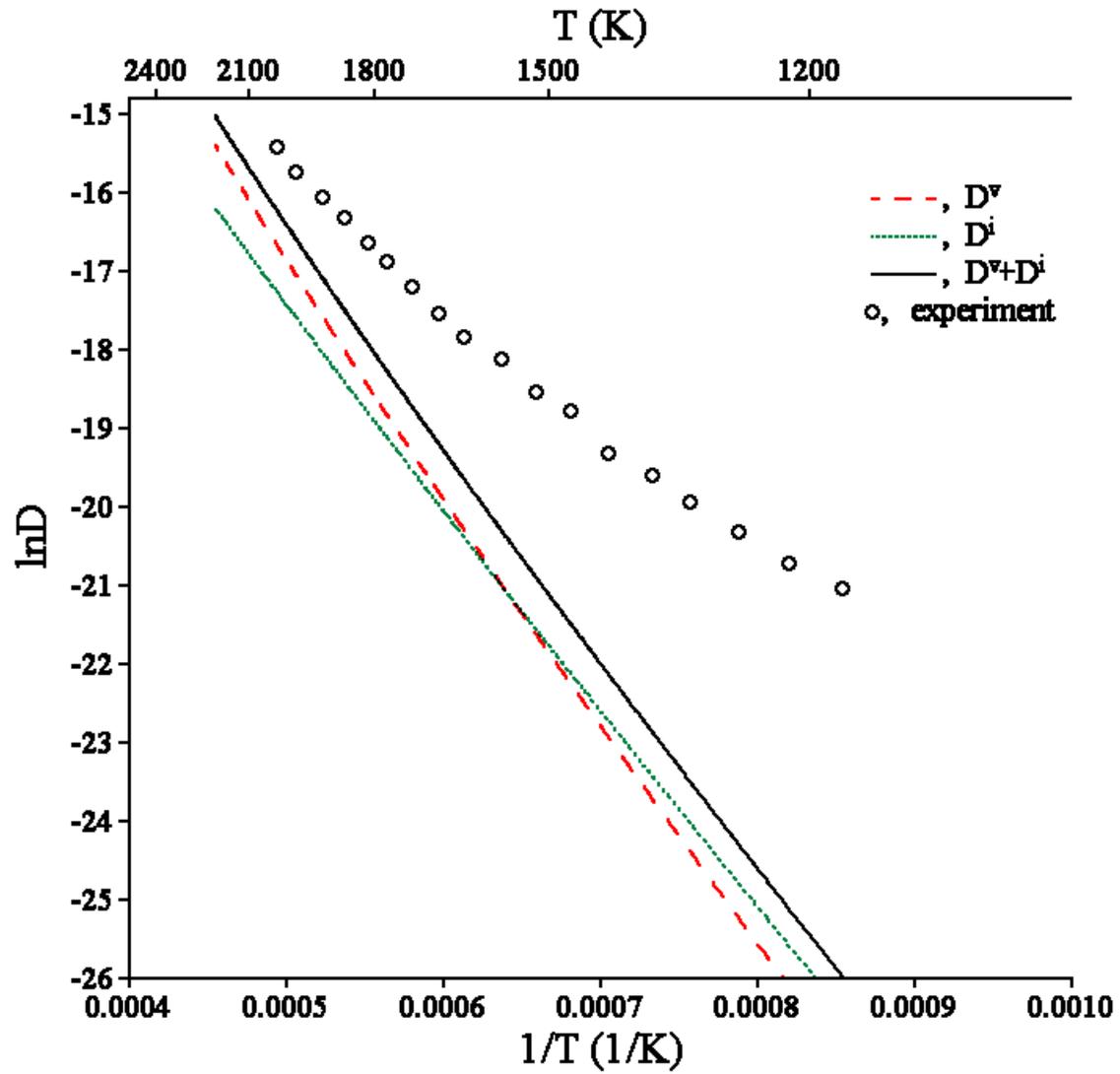

Figure 9. Diffusivity in bcc as function of temperature. The experimental data are from [15]